\documentclass[floats,aps,prl,eqnum,showpacs,twocolumn,amsmath]{revtex4}
\usepackage[dvips]{color,graphicx}
\usepackage{amsfonts}
\usepackage{amssymb}
\usepackage{latexsym}

\newcommand{\dalm}{\kern1pt\vbox{\hrule height 0.9pt\hbox{\vrule width
0.9pt\hskip 2.5pt\vbox{\vskip 5.5pt}\hskip 3pt\vrule width 0.3pt}\hrule height
0.3pt}\kern1pt}
\newcommand{\ma}[1]{\mbox{$\mathcal{#1}$}}

\newtheorem{The}{Theorem}

\begin{document}


\title{
Kaluza-Klein black hole with negatively curved extra dimensions in string generated gravity models}

\author{
$^{1,2,3}$Hideki Maeda\footnote{Electronic address:hideki@gravity.phys.waseda.ac.jp}
and
$^{4}$Naresh Dadhich\footnote{Electronic address:nkd@iucaa.ernet.in}}

\affiliation{
$^{1}$Graduate School of Science and Engineering, Waseda University, Tokyo 169-8555, Japan\\
$^{2}$Department of Physics, Rikkyo University, Tokyo 171-8501, Japan\\
$^{3}$Department of Physics, International Christian University, 3-10-2 Osawa, Mitaka-shi, Tokyo 181-8585, Japan\\
$^{4}$Inter-University Centre for Astronomy \& Astrophysics, Post Bag 4, Pune~411~007, India\\
}
\date{\today}

\begin{abstract}                
We obtain a new exact black-hole solution in Einstein-Gauss-Bonnet gravity  
with a cosmological constant which bears a specific relation to the Gauss-Bonnet coupling constant. 
The spacetime is a product of the usual $4$-dimensional manifold with a $(n-4)$-dimensional space of constant negative curvature, i.e., its topology is locally ${\ma M}^n \approx {\ma M}^4 \times {\ma H}^{n-4}$. 
The solution has two parameters and asymptotically approximates to the field of a charged black hole in anti-de~Sitter spacetime. 
The most interesting and remarkable feature is that the Gauss-Bonnet term acts like a Maxwell source 
for large $r$ while at the other end it regularizes the metric and weakens the central singularity. 
\end{abstract}


\maketitle

Superstring/M-theory is an attempt to unify all the forces in nature and for that it requires dimensions higher than usual four~\cite{superstring,Lukas}. 
In this theory, extra dimensions in our universe are considered to be compactified and hence are not accessible to present state of observations. 
There has emerged an attractive string-inspired braneworld model in which the universe we live in is a four-dimensional timelike hypersurface of a higher-dimensional bulk spacetime~\cite{large,Randall,DGP}. 
In the latter, the fundamental scale could be of order of TeV, and one of its consequences would be creation of tiny black holes, whose detection in the upcoming high energy collider becomes a distinct possibility~\cite{Bhformation}. 

Studies of higher dimensional spacetime and in particular black hole or black object have therefore been pursued vigorously and extensively. 
Recently, the black object called the black $p$-brane has been investigated for its stability which should be closely related to stability of the fundamental string theory object D-brane. 
The black $p$-brane is the $(n\ge 5)$-dimensional black object locally homeomorphic to ${\ma M}^{n-p} \times {\ma R}^{p}$, where ${\ma R}^{p}$ is the $p$-dimensional flat space.
The special case with $p=1$ is called the black string.
The instability of a black $p$-brane originally found by Gregory and Laflamme has occupied centrestage of investigations in this field~\cite{gl1993}. (See~\cite{kol2006} for a review.)

On the other hand, in the low-energy limit of heterotic superstring theory, the Gauss-Bonnet (GB) term naturally arises in the Lagrangian as the higher curvature correction to general relativity~\cite{Gross}. 
From a general standpoint it should also be included in the most general action for $n \ge 5$ which yields quasi-linear second order differential equation. 
There is also a purely classical motivation for higher dimensions based on the physical realization of dynamics of self interaction of gravity~\cite{dad}. 
For $n \ge 5$, the GB term must naturally be included along with Einstein-Hilbert action in the Lagrangian giving rise to Einstein-Gauss-Bonnet (E-GB) gravity.


The black $p$-brane solutions in E-GB gravity have recently been studied by several authors~\cite{got2006}. 
However, the generalization to the case with curved extra dimensions has not been done both in general relativity and in E-GB gravity. 
The purpose of this Letter is to report a new exact vacuum solution of E-GB gravity in a spacetime locally homeomorphic to ${\ma M}^{4} \times {\ma H}^{n-4}$ for $n \ge 6$, where ${\ma H}^{n-4}$ is the $(n-4)$-dimensional space of constant negative curvature.

We write action for $n\geq5$,
\begin{equation} 
\label{action}
S=\int d^nx\sqrt{-g}\biggl[\frac{1}{2\kappa_n^2}(R-2\Lambda+\alpha{L}_{GB}) \biggr]+S_{\rm matter},
\end{equation}
where $\alpha$ is the GB coupling constant and all other symbols having their usual meaning. 
The GB Lagrangian is the specific combination of Ricci scalar, Ricci and Riemann curvatures and it is given by 
\begin{equation}
{L}_{GB}=R^2-4R_{\mu\nu}R^{\mu\nu}+R_{\mu\nu\rho\sigma}R^{\mu\nu\rho\sigma}.
\end{equation}
This form of action follows from low-energy limit of heterotic superstring theory~\cite{Gross}. 
In that case, $\alpha$ is identified with the inverse string tension and is positive definite which is also required for stability of Minkowski spacetime. 
It should however be noted that it makes no contribution in the field equations for $n \le 4$.

The gravitational equation following from the action (\ref{action}) is given by 
\begin{equation}
{\ma G}^\mu_{~~\nu} \equiv {G}^\mu_{~~\nu} +\alpha {H}^\mu_{~~\nu} +\Lambda \delta^\mu_{~~\nu}=\kappa_n^2 {T}^\mu_{~~\nu}, \label{beq}
\end{equation}
where 
\begin{eqnarray}
{G}_{\mu\nu}&\equiv&R_{\mu\nu}-{1\over 2}g_{\mu\nu}R,\\
{H}_{\mu\nu}&\equiv&2\Bigl[RR_{\mu\nu}-2R_{\mu\alpha}R^\alpha_{~\nu}-2R^{\alpha\beta}R_{\mu\alpha\nu\beta}
\nonumber
\\
&& ~~~~
 +R_{\mu}^{~\alpha\beta\gamma}R_{\nu\alpha\beta\gamma}\Bigr]
-{1\over 2}g_{\mu\nu}{L}_{GB}.\label{def-H}
\end{eqnarray}

We consider the $n$-dimensional spacetime locally homeomorphic to ${\ma M}^{4} \times {\ma K}^{n-4}$ with the metric, $g_{\mu\nu}=\mbox{diag}(g_{AB},r_0^2\gamma_{ab})$, $A,B = 0, \cdots, 3;~a,b = 4, \cdots, n-1$. Here $g_{AB}$ is an arbitrary Lorentz metric on ${\ma M}^4$, $r_0$ is a constant and $\gamma_{ab}$ is the unit metric on the $(n-4)$-dimensional space of constant curvature ${\ma K}^{n-4}$ with its curvature ${\bar k} = \pm 1, 0$. 
Then ${\ma G}^\mu_{~~\nu}$ gets decomposed as follows:
\begin{eqnarray}
{\ma G}^A_{~~B}&=&\biggl[1+\frac{2{\bar k}\alpha(n-4)(n-5)}{r_0^2}\biggl]\overset{(4)}{G}{}^A_{~B} \nonumber \\
&&+\biggl[\Lambda-\frac{{\bar k}(n-4)(n-5)}{2r_0^2} \nonumber \\
&&-\frac{{\bar k}^2\alpha(n-4)(n-5)(n-6)(n-7)}{2r_0^4}\biggl] \delta^A_{~B},\label{dec1} \\
{\ma G}^a_{~~b}&=&\delta^a_{~~b}\biggl[-\frac12\overset{(4)}{R}+\Lambda-\frac{(n-5)(n-6){\bar k}}{2r_0^2} \nonumber \\
&&-\alpha\biggl\{\frac{{\bar k}(n-5)(n-6)}{r_0^2}\overset{(4)}{R}+\frac12 \overset{(4)}{L}_{GB} \nonumber \\
&&+\frac{(n-5)(n-6)(n-7)(n-8){\bar k}^2}{2r_0^4}\biggl\}\biggl],\label{dec2}
\end{eqnarray}
where the superscript $(4)$ means the geometrical quantity on ${\ma M}^4$.  

The decomposition immediately leads to a general result in terms of the following no-go theorem on ${\ma M}^4$: 
\begin{The} 
\label{the:1}
If (i) $r_0^2=-2{\bar k}\alpha(n-4)(n-5)$ and (ii) $\alpha\Lambda = -(n^2-5n-2)/[8(n-4)(n-5)]$, then  ${\ma G}^A_{~~B} = 0$ for $n \ge 6$ and ${\bar k}$ and $\Lambda$ being non-zero.
\end{The}
The proof simply follows from substitution of the conditions (i) and (ii) in  Eq.~(\ref{dec1}). 
As a corollary, it states that ${\ma M}^4$ cannot harbour any matter/energy distribution unless at least one of the conditions (i) and (ii) is violated. 

These conditions also imply for $\alpha > 0$, ${\bar k} = -1$ and $\Lambda < 0$. 
Hereafter we set ${\bar k} = -1$ and obtain the vacuum solution with $T_{\mu\nu} = 0$ satisfying the conditions (i) and (ii).
The governing equation is then a single scalar equation on ${\ma M}^4$, $\ma G^{a}_{~~b} = 0 $, which is given by
\begin{eqnarray}
\frac{1}{n-4}\overset{(4)}{R}+\frac{\alpha}{2} \overset{(4)}{L}_{GB}+\frac{2n-11}{\alpha(n-4)^2(n-5)}=0. \label{KKbasic}
\end{eqnarray}

We seek a static solution for a point mass with the metric on ${\ma M}^4$ reading as:
\begin{equation}
g_{AB}dx^Adx^B=-f(r)dt^2+\frac{1}{f(r)}dr^2+r^2d\Sigma_{2(k)}^2, \label{NdS2} 
\end{equation}
where $d\Sigma_{2(k)}^2$ is the unit metric on ${\ma K}^2$ and $k= \pm 1, 0$.
Then, Eq.~(\ref{KKbasic}) yields the general solution for the function $f(r)$: 
\begin{eqnarray}
f(r)&=&k+\frac{r^2}{2(n-4)\alpha}\biggl[1\mp\biggl\{1-\frac{2n-11}{3(n-5)} \nonumber \\
&&+\frac{4(n-4)^2\alpha^{3/2}\mu}{r^3}-\frac{4(n-4)^2\alpha^2q}{r^4}\biggl\}^{1/2}\biggl], \label{special} 
\end{eqnarray}
where $\mu$ and $q$ are arbitrary dimensionless constants. 
The solution does not have the general relativistic limit $\alpha \to 0$.
There are two branches of the solution indicated by sign in front of the square root in Eq.~(\ref{special}), which we call the minus- and plus-branches. 

The $n$-dimensional black hole with $(n-4)$-dimensional compact extra-dimensions is called the Kaluza-Klein black hole.
The warp-factor of the submanifold $r_0^2$ is proportional to GB parameter $\alpha$ which is supposed to be very small. 
Thus, compactifying ${\ma H}^{n-4}$ by appropriate identifications, we obtain the Kaluza-Klein black-hole spacetime with small and compact extra dimensions.
Here we shall mainly focus on the physical properties of the solution while a detailed study of its geometric structure and thermodynamical properties will be given in a forthcoming paper~\cite{next}.

The function $f(r)$ is expanded for $r \to \infty$ as
\begin{eqnarray}
f(r)&\approx& k\mp \frac{\alpha^{1/2} \mu\sqrt{3(n-4)(n-5)}}{r} \nonumber \\
&&\pm \frac{\alpha q\sqrt{3(n-4)(n-5)}}{r^2} \nonumber \\
&&+\frac{r^2}{2(n-4)\alpha}\left(1\mp\sqrt{\frac{n-4}{3(n-5)}}\right). \label{infty}
\end{eqnarray}
This is the same as the Reissner-Nordstr\"om-anti-de~Sitter (AdS) spacetime for $k=1$ in spite of the absence of the Maxwell field. This suggests that $\mu$ is 
the mass of the central object and $q$ is the charge-like new parameter. 

Further, the solution (\ref{special}) agrees with the solution in the Einstein-GB-Maxwell-$\Lambda$ system having the topology of ${\ma M}^n \approx {\ma M}^2 \times {\ma K}^{n-2}$ although it does not admit $n=4$.
The solution is given for $n \ge 5$ by 
\begin{equation}
ds^2=-g(r)dt^2+\frac{1}{g(r)}dr^2+r^2d\Sigma_{n-2(k)}^2 
\end{equation}
with
\begin{eqnarray}
g(r)&=&k+\frac{r^2}{2(n-3)(n-4)\alpha} \nonumber \\
&&\times \biggl[1\mp\biggl\{1+\frac{8(n-3)(n-4)\alpha\Lambda}{(n-1)(n-2)} \nonumber \\
&&+\frac{8(n-3)(n-4)\kappa_n^2\alpha M}{(n-2)V_{n-2}^{\bar k}r^{n-1}} \nonumber \\
&&-\frac{(n-4)\alpha\kappa_n^2 Q^2}{(n-2)\pi g_c^2r^{2(n-2)}}\biggl\}^{1/2}\biggr], \label{BDW}
\end{eqnarray}  
where $g_c$ is the coupling constant of the Maxwell field, and $M$ and $Q$ are mass and charge respectively~\cite{GBBH,tm2005b}.
$k$ is the curvature of ${\ma K}^{n-2}$ and a constant $V_{n-2}^k$ is its surface area on compactifications.
The non-zero component of the Maxwell field reads as 
\begin{eqnarray}
F_{rt}=\frac{Q}{r^{n-2}}
\end{eqnarray}  
representing the coulomb force of a central charge in $n$-dimensional spacetime.  

Thus the parameters $\mu$ and $q$ act as mass and ``charge'' respectively in spite of the absence of the Maxwell field. 
The new ``gravitational charge'' $q$ is generated by our choice of the topology of spacetime, splitting it into a product of the usual $4$-spacetime and a space of constant curvature. This splitting gives rise to the Kaluza-Klein modes 
which are known to generate such a gravitational charge known as the ``Weyl charge'' in the Randall-Sundrum braneworld model~\cite{sms2000}. There it is caused by the projection of the bulk Weyl curvature onto the brane, that is how it 
derives its name.

One of the first and the simplest black hole solutions on the brane obtained 
by Dadhich et al. by solving the gravitational equation on the brane is indeed given by the Reissner-Nordstr\"om metric~\cite{dad1}. (See~\cite{ag2005} for the rotating case.)
The Weyl charge was taken to be negative so as to work in unison with the mass. 
In our solution as well, $q$ must be negative so as to ward off any branch singularity indicated by vanishing of the expression under the square root in (\ref{special}). 
So we have a new Kaluza-Klein black hole with mass $\mu$ and Weyl charge 
$q < 0$ sitting in an AdS spacetime. 
It is really remarkable that our new solution asymptotically approximates (except for AdS background) to the brane black hole which was obtained in quite a different setting~\cite{dad1}. 
The common point between the two is splitting of spacetime into a bulk-brane system or a product. 
The Weyl charge seems to be caused by the Kaluza-Klein modes which require splitting of spacetime in some or the other way. Thus the Reissner-Nordstr\"om metric seems to be an asymptotically true description of black hole with GB adding AdS to it.

Clearly the global structure of our solution (\ref{special}) will be similar to that of the solution (\ref{BDW}) and it has been comprehensively studied in~\cite{tm2005b}.
Note that $f(0) = k\mp\sqrt{-q}$, which produces a solid angle deficit and it represents a spacetime of global monopole~\cite{bv}. 
This means that at $r = 0$ curvatures will diverge only as $1/r^2$ and so would be density which on integration over volume will go as $r$ and would therefore vanish. This indicates that singularity is weak as curvatures do not diverge strongly enough. 
 
\begin{figure}[tbp]
\includegraphics[width=.80\linewidth]{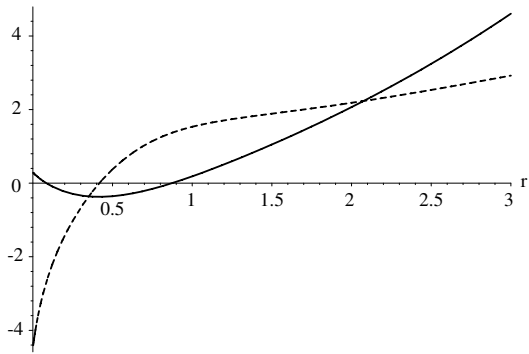}
\caption{
The function of $f(r)$ (solid line) and $df/dr$ (dashed line) in the minus-branch solution with $k=1$, $\alpha=0.1$, $n=6$, $\mu=2$, and $q=-0.5$.
The solution represents a black hole with inner and outer horizons.
There is a region with the repulsive gravity inside the outer horizon. 
}
\label{f1}
\end{figure}

\begin{figure}[tbp]
\includegraphics[width=.80\linewidth]{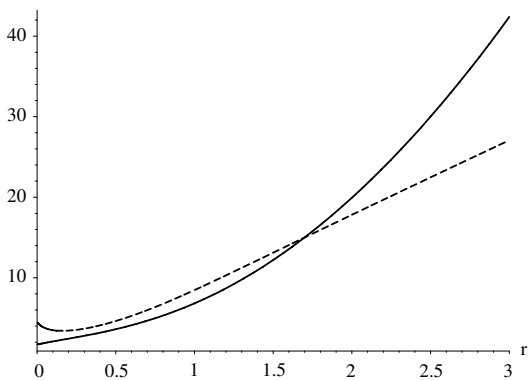}
\caption{
The function of $f(r)$ (solid line) and $df/dr$ (dashed line) in the plus-branch solution with $k=1$, $\alpha=0.1$, $n=6$, $\mu=2$, and $q=-0.5$.
The solution represents a naked singularity and the gravity is attractive everywhere. 
}
\label{f2}
\end{figure}

We plot $f(r)$ and $df/dr$ to get a good feel of the metric and 
gravitational field. Figs.~\ref{f1} and \ref{f2} respectively refer to black hole (in the minus-branch) and naked singularity (in the plus-branch) and they show that both metric and gravitational field always remain finite for finite $r$. 
That is why singularity is weak~\cite{tm2005b}.

The solution (\ref{special}) is the general solution of $\ma G^{a}_{~~b} = 0$ with the metric assumption (\ref{NdS2}) in addition to the conditions (i) and (ii).
Without the condition (i), it is obvious from Eq.~(\ref{dec1}) that the vacuum equations $\ma G^{A}_{~~B} = 0$ are identical to those in general relativity with a cosmological constant.
Then, by the generalized Birkhoff's theorem, the general solution of $\ma G^{A}_{~~B} = 0$ with the topology ${\ma M}^4 \approx {\ma M}^2 \times {\ma K}^{2}$ is Eq.~(\ref{NdS2}) with $f = k - \mu/r - \lambda r^2$ or the (anti-)Nariai solution.
Being confronted with $\ma G^{a}_{~~b} = 0$, the former will lead to $\mu=0$~\cite{next}.

Now arises the question of the interior solution. 
Our ansatz for local topology places a stringent constraint for interior of black hole. 
It is rather natural to consider the situation where $\alpha$ and $\Lambda$ in the interior are identical to those in the exterior.
Consequently, the condition (ii) holds in the interior, too.  
Then, by the contraposition of the no-go theorem proven above, the matter interior represented by the metric $g_{\mu\nu}=\mbox{diag}(g_{AB},r_0^2\gamma_{ab})$ cannot satisfy the condition (i).
Therefore, such an interior solution cannot be attached to our vacuum solution.
However, it could be attached to the interior with the metric $g_{\mu\nu}=\mbox{diag}(g_{AB},S(x^D)^2\gamma_{ab})$, where $S$ is a scalar on ${\ma M}^4$, at $S^2=r_0^2=2\alpha(n-4)(n-5)$.
The matching problem to the interior is a very involved and difficult problem which we shall address in our future studies. 
   
We have thus found a new Kaluza-Klein black hole solution of Einstein-Gauss-Bonnet gravity with topology of product of the usual $4$-spacetime with a negative constant curvature space. 
In this solution we have brought the GB effects down on four dimensional black hole as envisaged in~\cite{dad}.
Asymptotically it resembles a charged black hole in AdS background while at the other end it approximates to a global monopole. 
What really happens is that GB term regularizes the metric and weakens the singularity while the presence of extra dimensional hyperboloid space generates the Kaluza-Klein modes giving rise to the Weyl charge. 
This is indeed the most interesting and remarkable feature of the new solution which needs to be probed further for greater insight and application~\cite{next}. 


\acknowledgments
The authors would like to thank Reza Tavakol and Umpei Miyamoto for discussions.
HM would like to thank Hideki Ishihara and Takashi Torii for useful comments. 
HM would also to thank IUCAA for warm hospitality where the work was conceived and 
formulated. 
HM was supported by the Grant-in-Aid for Scientific
Research Fund of the Ministry of Education, Culture, Sports, Science
and Technology, Japan (Young Scientists (B) 18740162).




\begin{references}
\bibitem{superstring}
J.~Polchinski, 
{\it String Theory: An Introduction to the Bosonic String}
(Cambridge University Press, Cambridge, England, 1998);
{\it String Theory: Superstring Theory and Beyond}
(Cambridge University Press, Cambridge, England, 1998).
\bibitem{Lukas} 
   P. Ho\v rava and E. Witten, 
   Nucl. Phys. {\bf B475}, 94 (1996);
   A. Lukas, B. A. Ovrut, K.S. Stelle, and D. Waldram,
   Phys. Rev. D {\bf 59}, 086001 (1999);
   A. Lukas, B. A. Ovrut, and D. Waldram,
   Phys. Rev. D {\bf 60}, 086001 (1999).
\bibitem{large} 
   N. Arkani-Hamed, S. Dimopoulos, and G. Dvali, 
   Phys. Lett. {\bf B429}, 263 (1998);
   I. Antoniadis, N. Arkani-Hamed, S. Dimopoulos, and G. Dvali,
   Phys. Lett. {\bf B436}, 257 (1998).
\bibitem{Randall} 
   L. Randall and R. Sundrum,
   Phys. Rev. Lett. {\bf 83}, 3370 (1999);
   Phys. Rev. Lett. {\bf 83}, 4690 (1999).
\bibitem{DGP} 
   G. Dvali, G. Gabadadze, and M. Porrati, 
   Phys. Lett. {\bf B485}, 208 (2000);
   G. Dvali and G. Gabadadze, 
   Phys. Rev. D {\bf 63}, 065007 (2001);
   G. Dvali, G. Gabadadze, and M. Shifman, 
   Phys. Rev. D {\bf 67}, 044020 (2003).
\bibitem{Bhformation} 
   S. Dimopoulos, G. Landsberg, 
   Phys. Rev. Lett. {\bf 87} 161602 (2001);
   A. Chamblin and G.C. Nayak, 
   Phys. Rev. D {\bf 66} 091901 (2002);
   S. B. Giddings and S. Thomas, 
   Phys. Rev. D {\bf 65} 056010 (2002).
\bibitem{gl1993}
R.~Gregory and R.~Laflamme,
Phys. Rev. Lett. {\bf 70}, 2837 (1993);
Nucl. Phys. {\bf B428}, 399 (1994).
\bibitem{kol2006}
B.~Kol,
Phys. Rep. {\bf 422}, 119 (2006).
\bibitem{Gross} 
D.J.~Gross and J.H.~Sloan, 
Nucl. Phys. {\bf B291}, 41 (1987);
M.C.~Bento and O.~Bertolami,
Phys. Lett. {\bf B368}, 198 (1996).
\bibitem{dad} 
   N.~Dadhich, 
   hep-th/0509126.
\bibitem{got2006} 
T.~Kobayashi and T.~Tanaka,
Phys. Rev. D {\bf 71}, 084005 (2005); 
C.~Sahabandu, P.~Suranyi, C.~Vaz, and L.C.R.~Wijewardhana,
Phys. Rev. D {\bf 73}, 044009 (2006); 
D.~Kastor and R.~Mann, 
JHEP {\bf 0604}, 048 (2006);
G.~Giribet, J.~Oliva, and R.~Troncoso, 
JHEP {\bf 0605}, 007 (2006). 
\bibitem{next} 
   H.~Maeda and N.~Dadhich, 
   (to be published).
\bibitem{GBBH} 
   R. -G. Cai, 
   Phys. Rev. D {\bf 65}, 084014 (2002);
   D. G. Boulware and S. Deser, 
   Phys. Rev. Lett. {\bf 55}, 2656 (1985);
   J. T. Wheeler,
   Nucl. Phys. {\bf B268}, 737 (1986);
   D. L.~Wiltshire,
   Phys. Lett. {\bf B169}, 36 (1986);
   Phys. Rev. D {\bf 38}, 2445 (1988).
\bibitem{tm2005b} 
T.~Torii and H.~Maeda, 
Phys. Rev. D {\bf 71}, 124002 (2005);
Phys. Rev. D {\bf 72}, 064007 (2005).
\bibitem{sms2000} 
T.~Shiromizu, K-i.~Maeda, and M.~Sasaki, 
Phys. Rev. D {\bf 62}, 024012 (2000).
\bibitem{dad1} 
   N.~Dadhich, R.~Maartens, P.~Papadopoulos, and V.~Rezania, 
   Phys. Lett. {\bf B487}, 1 (2000).
\bibitem{ag2005} 
A.N.~Aliev and A.E.~Gumrukcuoglu,
Phys. Rev. D {\bf 71}, 104027 (2005).
\bibitem{bv} 
   M.~Barriola and A.~Vilenkin, 
   Phys. Rev. Lett. {\bf 63}, 341 (1989);
   N.~Dadhich, K.~Narayan, and U.~Yajnik, 
   Pramana {\bf 50}, 307 (1998).










\end{references}
\end{document}